\RequirePackage{fix-cm}
\documentclass[twocolumn,epjc3]{svjour3}  
\smartqed  
\RequirePackage{graphicx}

\usepackage{url}
\usepackage{graphicx}

\usepackage{latexsym}
\usepackage{amssymb}
\usepackage{amsfonts} 
\usepackage{amsmath}

\newcommand{\bea}{\begin{eqnarray}}
\newcommand{\eea}{\end{eqnarray}}
\newcommand{\be}{\begin{equation}}
\newcommand{\ee}{\end{equation}}

\newcommand{\dd}{{\rm d}}
\newcommand{\nn}{\nonumber}

\newcommand{\cmnt}[1]{}
\newcommand{\insl}{\not\!\, \in}

\newcommand{\gm}{\gamma}
\newcommand{\Gm}{\Gamma}
\newcommand{\dl}{\delta}

\newcommand{\ep}{\varepsilon}

\newcommand{\lm}{\lambda}

\journalname{Eur. Phys. J. C}
\begin{document}

\title{On the status of expansion by regions
}


\author{Tatiana Yu.\ Semenova\thanksref{e1,addr1}
        \and
        Alexander V.\ Smirnov\thanksref{e2,addr2} 
        \and
        Vladimir A.\ Smirnov\thanksref{e3,addr3}
}

\thankstext{e1}{e-mail: station@list.ru}
\thankstext{e2}{e-mail: asmirnov80@gmail.com}
\thankstext{e3}{e-mail: smirnov@theory.sinp.msu.ru}


\institute{Department of Mechanics and Mathematics, Moscow State University, 
119992 Moscow, Russia \label{addr1}
           \and
           Research Computing Center, Moscow State University, 
119992 Moscow, Russia \label{addr2}
           \and
           Skobeltsyn Institute of Nuclear Physics of Moscow State 
University, 119992 Moscow, Russia  \label{addr3}
}

\date{Received: date / Accepted: date}

\maketitle

\begin{abstract}
We discuss the status of expansion by regions, i.e. a well-known strategy to obtain 
an expansion of a given
multiloop Feynman integral in a given limit where some kinematic invariants and/or masses
have certain scaling measured in powers of a given small parameter. Using the Lee-Pomeransky parametric representation,
we formulate the corresponding prescriptions in a
simple geometrical language and make a conjecture that they hold even in a much more general case. 
We prove this conjecture 
in some partial cases. 
\end{abstract}
 
\section{Introduction}
 
If a given Feynman integral depends on kinematic invariants and masses which essentially differ in scale, 
a very natural and often used idea is to expand it in powers of a given small parameter. As a result, the integral 
can be written as a series of factorized quantities which are simpler than the original integral itself and it can be 
substituted by a sufficiently large number of terms of such an expansion.  
The strategy of expansion by regions~\cite{Beneke:1997zp} (see also \cite{Smirnov:2002pj} and Chapter~9 of \cite{Smirnov:2012gma})
introduced and applied in the case
of threshold expansion~\cite{Beneke:1997zp} is a strategy to obtain an expansion of a given
multiloop Feynman integral in a given limit specified by scalings of kinematic invariants and/or masses
characterized by powers of a given small parameter of expansion.
For example, for a limit with two variables, $q^2$ and $m^2$, where $m^2\ll q^2$ and the parameter of expansion
is $m^2/q^2$, one analyzes various regions in a given integral over loop momenta 
and, in every region, expands the integrand, i.e. a product of propagators,
in parameters which are there small. Then the integration in the integral with so expanded 
propagators is extended to the whole domain of the loop momenta and, finally, one obtains
an expansion of the given integral as the corresponding sum over the regions.

Although this strategy certainly looks suspicious for mathematicians it was successfully applied in
numerous calculations. It has the status of experimental mathematics and should be applied with care,
starting, first, from one-loop examples, by checking results by independent methods.
Starting from the analysis in a toy example of an expansion of a one-dimensional 
integral~\cite{Beneke} presented in \cite{Smirnov:2002pj},
Jantzen \cite{Jantzen:2011nz} provided detailed explanations of how this strategy works by
starting from regions determined by some inequalities and covering the whole integration space
of the loop momenta, then expanding the integrand and then extending integration and
analyzing all the pieces which are obtained, with the hope that `readers would be convinced that 
the expansion by regions is a well-founded method'. However, this interesting and instructive
analysis can hardly be considered as a base of mathematical proofs. Let us realize that 
we are dealing with dimensionally regularized Feynman integrals, i.e. integrals
over loop momenta of space-time dimension $d=4-2\ep$ which is considered as
a complex regularization parameter. Therefore it is not clear
in which sense inequalities and limits for these integrals are understood because
the integrands and the integrals are functions of $d$-dimensional loop and/or external momenta so that they should be
treated like some algebraic objects rather than usual functions in integer numbers of dimensions.
In practice, one usually does not bother about such problems and performs calculations implicitly
applying some axioms for the integration procedure, and a consistency of the whole calculation
checked in some way looks quite sufficient.
 
A well-known way to deal with dimensionally regularized multiloop Feynman integrals is
to use, for a given graph, the corresponding Feynman parametric representation which up to an overall gamma function and a power
of $(i\pi^{d/2})$ (which we will always omit) takes the following form in the case of all powers $a_i$ of propagators
$1/(-p^2+m_l^2-i0)^{a_i}$
equal to one:
\bea
\int_0^\infty \ldots\int_0^\infty \,
U^{n-(h+1) d/2} F^{h d/2-n}
&& \nn \\ &&  \hspace*{-35mm}
\times\delta\left( \sum x_i-1\right) \,
 \dd x_1 \ldots \dd x_n\;,
\label{alpha-d-mod}
\eea
where $n$ is the number of lines (edges), $h$ is the number of loops (independent circuits) of the graph,
\begin{equation}
F= -V +U\sum m^2_l x_l \;,
\label{Wfunction}
\end{equation}
and $U$ and $V$ are two basic functions (Symanzik polynomials, or graph polynomials)
\bea
U& = & \sum_{T\in T^1} \prod_{l\insl  T} x_l
 \;,
\label{Dform}  \\
V&=& \sum_{T\in T^2}  
\prod_{l\insl  T} x_l
\left( q^T\right)^2
 \; .
\label{Aform}
\eea
In (\ref{Dform}), the sum runs over
trees
of the given graph,
and, in (\ref{Aform}),
over {\em 2-trees}, i.e. subgraphs that do not involve loops
and consist of two connectivity components; $\pm q^T$ is
the  sum of the external momenta that flow
into one of the connectivity components of the 2-tree $T$. 
The products of the Feynman parameters involved are taken
over the lines that do not belong to a given tree or a 2-tree $T$.
As is well known, one can choose the sum in the argument of the delta-function
over any subset of lines. In particular, one can choose just one Feynman parameter,
$x_l$, and then the integration will be over the other parameters at $x_l=1$.
The functions $U$ and $V$ are homogeneous with respect to Feynman parameters, 
with the homogeneity degrees $h$ and
$h+1$, respectively.

The parametric representation in the case where propagators enter with general powers $a_i$ 
can be obtained from (\ref{alpha-d-mod}) by including the overall factor 
$\Gamma(a-h d/2)/(\prod_i \Gamma(a_i))$,
with $a=\sum a_i$,
and the product $\prod_i x_l^{a_i-1}$ in the integrand.
The representation with negative integer indices $a_i=-n_i$ can be obtained from this one
by taking the limit $a_i \to -n_i$ where the pole at $a_i = -n_i$ arising from $x_i^{a_i-1}$
is cancelled by the pole of $\Gamma(a_i)$ in the denominator.
However, to make the presentation simpler, we will consider only the case of all the indices equal to one.

The expansion by regions was also formulated in the language of the corresponding parametric 
integrals~\cite{Smirnov:1999bza} (see also \cite{Smirnov:2002pj} and Chapter~9 of \cite{Smirnov:2012gma}). 
One can consider quite general limits for a Feynman integral which depends on external momenta $q_i$
and masses and is a scalar function of kinematic invariants $q_i \cdot q_j$ and squares of masses
and assume that each kinematic invariant and a mass squared has certain scaling
$\rho^{\kappa_i}$ where $\rho$ is a small parameter.
A non-trivial point when applying the strategy of expansion by regions,
either in momentum space or in parametric representation,
is to understand which regions are relevant to a given limit.
For example, for the threshold expansion, these are hard, potential, soft and ultrasoft regions,
as it was claimed in~\cite{Beneke:1997zp} and further confirmed in practice in multiple calculations.

A systematical procedure to find relevant regions was developed in Ref.~\cite{Pak:2010pt}
using Feynman parametric representation (\ref{alpha-d-mod}) and geometry of polytopes connected with the basic  
functions $U$ and $F$. This procedure was implemented as a public computer code {\tt asy.m}~\cite{Pak:2010pt}
which is now included in the code {\tt FIESTA} \cite{Smirnov:2015mct}. Using this code one can 
not only find relevant regions but also obtain the corresponding terms of expansion
and evaluate numerically coefficients at powers and logarithms of the given expansion
parameter. Although there is no mathematical justification of this procedure, numerous applications
have shown that the code {\tt asy.m} works consistently at least in the case where
all the terms in the function $F$ are positive. An attempt to extend this procedure and
the corresponding code {\tt asy.m} to some cases where some terms of the function $F$ are negative was made
in Ref.~\cite{Jantzen:2012mw} where it was explained how potential and Glauber regions
can be revealed.\footnote{After our paper has been sent to the archive, a new approach 
(based on Landau equations) to reveal regions corresponding to a given limit has appeared~\cite{Ananthanarayan:2018tog}. 
Its authors show on examples that the potential and Glauber regions can be revealed within their prescriptions.}
 
We find it very natural to use Feynman parametric representations and the geometrical description
of expansion introduced in Ref.~\cite{Pak:2010pt} to mathematically prove expansion by regions.
In fact, for the moment, only an indirect proof of expansion by regions, for limits typical of Euclidean space
(where one has two different regions which can be called large and small) exists, --
see the proof for the off-shell large-momentum limit in \cite{Smirnov:1990rz} and Appendix~B.2 of \cite{Smirnov:2002pj}.
The point is that, for limits typical of Euclidean space (for example, the off-shell large-momentum limit or 
the large-mass limit), one can write down the corresponding expansion in terms 
of a sum over certain subgraphs of a given graph \cite{Chetyrkin:1988zz,Chetyrkin:1988cu,Gorishnii:1989dd},
and there is a correspondence between these subgraphs and
their loop momenta which are considered large while the other loop momenta are considered small.

We would like to emphasize that in order to try to mathematically prove expansion by regions,
it looks preferable and mathematically natural to use a recently suggested representation by Lee and 
Pomeransky (LP) \cite{Lee:2013hzt} instead of the well-known representation (\ref{alpha-d-mod}).
Up to an overall product of gamma functions,
this representation has the form
\bea
G(\ep)=
\int_0^\infty \ldots\int_0^\infty \,
P^{-\dl}  \dd x_1 \ldots \dd x_n\;,
\label{L-P}
\eea
where $\dl=2-\ep$ and
\bea
P= U +F \;.
\label{Pol}
\eea
One can obtain (\ref{alpha-d-mod}) from (\ref{L-P}) by \cite{Lee:2013hzt}
inserting  the relation $1 = \int \dl(\sum_i x_i -\eta)\dd \eta$,  scaling $x \to \eta x$ 
and integrating over $\eta$.

We believe that the prescriptions of expansion by regions hold also for integrals (\ref{L-P})
with a general polynomial $P$ with positive coefficients and not only for polynomials
of the form (\ref{Pol}) where the two terms are basic functions for some graph.
The goal of our paper is, at least, to formulate prescriptions of expansion by regions for general polynomials
with positive coefficients in an unambiguous mathematical language, 
to justify how terms of the leading order of expansion are constructed
and to draw attention of both physicists and mathematicians 
who might find it interesting to prove it in a general order of expansion.

In the next section we use the geometrical description of expansion by regions on which
the code {\tt asy.m}~\cite{Pak:2010pt} was based.
In this paper, we consider limits with two scales where one introduces a small parameter as their ratio.
Let us emphasize that this can be various important limits which are typical of Minkowski space,
for example, the Sudakov limit or the Regge limit (with $|t| \ll |s|$ where $s$ and $t$ are
Mandelstam variables.)
In this description, regions correspond to special facets of the Newton polytope
associated with the product of $U F$ of the two basic polynomials in (\ref{alpha-d-mod}).
We immediately switch here to prescriptions based on the LP~\cite{Lee:2013hzt}
parametric representation~(\ref{L-P}) and
formulate prescriptions for a general polynomial with positive coefficients, rather than polynomial (\ref{Pol}).
Therefore, these prescriptions will be based on facets of the corresponding Newton polytope.
Of course, prescriptions based on representation (\ref{L-P}) are algorithmically preferable because
the degree of the sum of the two basic polynomials is smaller than the degree of their product
$U F$ (used in {\tt asy.m}) so that looking for facets of the corresponding Newton polytope
becomes a simpler procedure\footnote{In fact, this step is performed within {\tt asy.m} with the help
of another code {\tt qhull}. It is most time-consuming and can become problematic in higher-loop
calculations.}. Therefore, the current version of the code {\tt asy.m} included in {\tt FIESTA} \cite{Smirnov:2015mct}
is now based on this more effective procedure.

Since we are oriented at mathematical proofs we want to be mathematically correct.
Let us realize that up to now we did not discuss whether integral (\ref{alpha-d-mod}) or (\ref{L-P}) 
can be understood as a convergent integral at some values of $d$. 
Let us keep in mind a situation where a Feynman integral is both ultravioletly and infrared divergent
so that increasing Re$(\ep)$ regulates ultraviolet divergences and decreasing Re$(\ep)$ regulates infrared divergences.
Such situations are not exotic at all.
However, in practical calculations
of Feynman integrals one usually does not bother about the existence of such a convergence domain
and/or tries to define the given integral in some other way if such a domain does not exist.
Well, after calculation are made, one has a result which is a function of $d=4=2\ep$ usually
presented by first terms of a Laurent expansion near $\ep=0$, and such a result is well defined!

In Section~3, we refer to some papers where attempts to define Feynman integrals {\em before} calculations are made
and comment on how Feynman integrals are understood when they are evaluated.
Then we turn to parametric representation~(\ref{L-P}) in Section~4 and explain how we can define this representation
in terms of convergent integrals.
In Section~5, we explicitly show that, in the case of Feynman integrals, i.e. where the polynomial is given by (\ref{Pol}),
with two basic functions constructed for a given graph, the two kinds of prescriptions based either on the
Feynman parametric representation or on the LP parametric representation
are equivalent.

Equipped with our definition based on analytic regularization, we then turn in Section~6 to the main conjecture,
prove it it in the leading order in a special situation and 
analyze it in the leading order of expansion in the general situation. 
In Section~7, we prove the main conjecture in the simple case, where only one facet contributes.
In Section~8, we summarize our results and discuss perspectives.

\section{The main conjecture}
  
Let us formulate the  main conjecture about expansion by regions for integral (\ref{L-P})
with a polynomial with positive coefficients in the case of
limits with two kinematic invariants and/or masses of essentially different scale, where one introduces one parameter, $t$,
which is the ratio of two scales and is considered small. 
Then the polynomial in Eq.~(\ref{L-P}) is a function of Feynman parameters and $t$,
\bea
P(x_1,\ldots,x_n,t)=\sum_{w\in S} c_{w}x_1^{w_1}\ldots x_n^{w_n} t^{w_{n+1}}\;,
\label{polynomial}
\eea
where $S$ is a finite set of points $w=(w_1,...,w_{n+1})$ and $c_{w}>0$. 
The Newton polytope $\mathcal{N}_P$ of $P$ 
is the convex hull of the points $w$ in the $n+1$-dimensional Euclidean space $\mathbb{R}^{n+1}$
equipped with the scalar product $v\cdot w=\sum_{i=1}^{n+1} v_i w_i$. 
A facet of $\mathcal{N}_P$ is a face of maximal dimension, i.e. $n$.
  
{\bf The main conjecture.} 
The asymptotic expansion of (\ref{L-P}) in the limit  
$t\to +0$ is given by  
\bea
G(t,\ep)\sim \sum_{\gm}
\int_0^\infty\ldots\int_0^\infty
\left[
M_{\gm} \left(P(x_1,\ldots,x_n,t)\right)^{-\dl}
\right]
&& \nn \\ &&  \hspace*{-42mm}
\times
 \dd x_1 \ldots \dd x_{n} \,,
\label{rhs}
 \eea
where the sum runs over facets of the Newton polytope $\mathcal{N}_P$ 
for which the normal vectors
$r^{\gm} = (r^{\gm}_1,\ldots,r_{n+1}^{\gm})$ oriented inside the polytope
have $r_{n+1}^{\gm}>0$. Let us normalize these vectors by $r_{n+1}^{\gm}=1$
and let us call such facets {\em essential}. To describe operators $M_{\gm}$ we need, first, to introduce 
some notation and a number of definitions.

The symbol $\sim$ in (\ref{rhs}) is the standard symbol of an asymptotic expansion. As it will be explained shortly,
every term in the right-hand side of (\ref{rhs}) is homogeneous with respect to the expansion parameter, $t$, so that one can sort out 
various terms of the expansion according to their order in $t$ and construct the sum of first terms, up to order $t^N$.
Then, according to the definition of the asymptotic expansion, the corresponding remainder defined as the difference
between the initial integral and these first terms, is of order $o(t^n)$.

The contribution of a given essential facet is defined by the change of variables $x_i \to  t^{r^{\gm}_i} x_i$
in the integral (\ref{L-P}) and expanding the resulting integrand in powers of $t$.
This leads to the following definitions.

For a given essential facet ${\gm}$, let us define the polynomial
\bea
P^{\gm}(x_1,\ldots,x_n,t)&=&P(t^{r^{\gm}_{1}} x_1,\ldots,t^{r^{\gm}_{n}} x_n,t)
\nn \\
&\equiv& 
\sum_{w\in S} c_{w} x_1^{w_1}\ldots x_n^{w_n} t^{w\cdot r^{\gm}}   \;.
\label{Pgm}
\eea
The scalar product $w\cdot r^{\gm}$ is proportional to the projection of the point  $w$ on the vector $r^{\gm}$.
For $w\in S$, it takes a minimal value for all the points belonging to the considered facet $w\in S \cap \gm$.
Let us denote it by $L(\gm)$.

The polynomial (\ref{Pgm}) can be represented as
\bea
t^{L(\gm)}\left(
P^{\gm}_0(x_1,\ldots,x_n) + P^{\gm}_1(x_1,\ldots,x_n,t)\right)
\label{Pgm1}\;,
\eea
where
\bea
P^{\gm}_0(x_1,\ldots,x_n) &=&\sum_{w\in S\cap \gm} c_{w} x_1^{w_1}\ldots x_n^{w_n}\,,
\label{P1}
\\
P^{\gm}_1(x_1,\ldots,x_n,t)&=&\sum_{w\in S\setminus \gm} 
c_{w} x_1^{w_1}\ldots x_n^{w_n} t^{w\cdot r^{\gm}-L(\gm)} 
\label{P2}\;.
\eea
The polynomial $P^{\gm}_0$ is independent of $t$ while $P^{\gm}_1$ can be represented as a linear combination 
of positive rational powers of $t$ with coefficients which are polynomials of $x$.

For a given facet $\gm$, let us define the operator  
\bea
M_{\gm} \left(P(x_1,\ldots,x_n,t)\right)^{-\dl} 
= t^{\sum_{i=1}^n r^{\gm}_i-L(\gm)\dl}
&& \nn \\ &&  \hspace*{-55mm}
\times {\cal T}_{t}
\left( P^{\gm}_0(x_1,\ldots,x_n) +  P^{\gm}_1(x_1,\ldots,x_n,t) 
\right)^{-\dl}
\nn \\ && \hspace*{-55mm} 
\nn
= t^{\sum_{i=1}^n r^{\gm}_i-L(\gm)\dl}\ \left( P^{\gm}_0(x_1,\ldots,x_n)\right)^{-\dl}+\ldots 
\label{M-operator}
\eea
where ${\cal T}_{t}$ 
performs an asymptotic expansion in powers of
$t$ at  $t= 0$.
 
{\bf Comments.}  
\begin{itemize}
\item 
An operator $M_{\gm}$ can equivalently be defined by introducing a parameter $\rho_{\gm}$,
replacing $x_i$ by $\rho^{r^{\gm}_{i}} x_i$ , pulling an overall power of $\rho_{\gm}$,
expanding in $\rho_{\gm}$ and setting $\rho_{\gm}=1$ in the end. It is reasonable to use this variant 
when one needs to deal with products of several operators $M_{\gm}$.
\item 
The leading order term of a given facet $\gm$ corresponds to the leading order of the operator
$M^0_{\gm} $:
\bea
&& \int_0^\infty\ldots\int_0^\infty
\left[
M^0_{\gm} \left(P(x_1,\ldots,x_n,t)\right)^{-\dl}
\right]
 \dd x_1 \ldots \dd x_{n} 
 \nn \\
&=&t^{-L(\gm)\dl+\sum_{i=1}^n r^{\gm}_i}
\nn \\
&& \hspace*{-3mm} \times
\int_0^\infty \ldots\int_0^\infty \,
\left(P^{\gm}_0(x_1,\ldots,x_n)\right)^{-\dl}  \dd x_1 \ldots \dd x_n\;.
\label{L-P-LO}
\eea
\item
In fact, with the above definitions, we can write down the equation of the hyperplane generated by a given facet $\gm$
as follows
\bea
w_{n+1}=-\sum_{i=1}^n r^{\gm}_i w_i+L(\gm)\;. 
\label{hyperplane_gamma}
\eea
\item
Let us agree that the action of an operator $M_\gm$ on an integral reduces to the action of
$M_\gm$ on the integrand described above.
Then we can write down the expansion in a shorter way,
\bea
G(t,\ep)\sim \sum_{\gm} M_{\gm}  G(t,\ep)
\label{rhs-ec}
 \eea
\item
In the usual Feynman parametrization (\ref{alpha-d-mod}),
the expansion by regions in terms of operators $M_{\gm}$ is formulated in a similar way, and this is exactly
how it is implemented in the code {\tt asy.m}~\cite{Pak:2010pt}. The expansion can be written in the same form
(\ref{rhs-ec}) but the operators $M_{\gm}$ act on the product of the two basic polynomials 
$U$ and $F$ raised to certain powers present in (\ref{alpha-d-mod}). Now, each of the two polynomials is decomposed
in the form (\ref{Pgm1}) and so on.
\item
It is well known that dimensional regularization might be not sufficient to regularize individual contributions
to the asymptotic expansion. A natural way to overcome this problem is to introduce an auxiliary
analytic regularization, i.e. to introduce additional exponents $\lm_i$ to power of the propagators.
This possibility exists in the code {\tt asy.m}~\cite{Pak:2010pt}
included in {\tt FIESTA} \cite{Smirnov:2015mct}. One can choose these additional parameters in some way
and obtain a result in terms of an expansion in $\lm_i$ followed by an expansion in $\ep$.
If an initial integral can be well defined as a function of $\ep$ then
the cancellation of poles in $\lm_i$ serves as a good check of the calculational procedure, so that
in the end one obtains a result in terms of a Laurent expansion in $\ep$ up to a desired order.
We will systematically exploit analytic regularization below for various reasons.
\item
We consider the case of two kinematic parameters for simplicity.
In the general case, with several kinematic invariants $q_i \cdot q_j$ and squares of masses,
where each of these variables, $s_i$, has certain scaling, i.e. $s_i \to \rho^{\kappa_i}s_i$,
with $\rho$ a small parameter,
one can formulate similar prescriptions. Then, the expansion is given by
a similar sum over facets of a Newton polynomial which is determined for each choice
of the variables $s_i$. (This is how the code {\tt asy.m} ~\cite{Pak:2010pt} works in this case.) 
\end{itemize}

\section{Convergence and sector decompositions}

When formulating the main conjecture in the previous section we did not discuss conditions under which
integral~(\ref{L-P}) is convergent. As is well known, dimensional regularization is introduced
for Feynman integrals, i.e. when polynomial is given by (\ref{Pol}), in order that various divergences
become regularized so that the integral becomes a meromorphic function of the regularization parameter
$\ep$. Then one can deal with the regularized quantity where divergences manifest themselves
as various poles at $\ep=0$. However, a given Feynman integral can have both ultraviolet divergences
which can be regularized by increasing Re$(\ep)$ and (off-shell or on-shell) 
infrared\footnote{We follow the terminology introduced in the sixties and seventies.
Ultraviolet (infrared) divergences arise from integration over large (small) 
loop momenta. By off-shell infrared divergences we mean divergences at small 
loop momenta in situations where external momenta are not put on a mass shell.
In particular, external momenta can be considered Euclidean (any partial sum
of external momenta is space-like) (see, for example, Ref. \cite{Speer:1975dc}), or
a Feynman integral can be considered as a tempered distribution with respect
to external momenta (for example, in very well-known papers on renormalization 
\cite{Hepp:1966eg,Breitenlohner:1977hr,Breitenlohner:1976te1,Breitenlohner:1976te2}). 
On-shell infrared divergences appear when an external momentum is considered
on a mass shell, $p^2=m^2$, in particular a massless mass shell. In the latter case,
collinear divergences can appear due to integration near light-like lines.}
as well as collinear
divergences which can be regularized by decreasing Re$(\ep)$. Then, typically, there is no domain of
$\ep$ where the integral is convergent. In numerous calculations, one does not bother about this problem.
Rather, various methods of evaluating Feynman integrals are applied and in the end of a calculation,
one arrives at a result which looks like several terms of a Laurent expansion in $\ep$.
 
Let us now remember that there is a mathematical definition of a dimensionally regularized Feynman integral 
in the case where both ultraviolet and off-shell infrared divergences are present.
Speer defines \cite{Speer:1975dc} such an integral\footnote{without massless detachable subgraphs; this means that there are
no one-vertex-irreducible subgraphs with zero incoming momenta. The corresponding integrals would be
scaleless integrals.} as an analytic continuation of the corresponding dimensionally and analytically regularized
integral, i.e. with all propagators $1/(-p_l^2+m^2-i 0)$ replaced by $1/(-p_l^2+m^2-i 0)^{1+\lm_l}$,
from a domain of analytic regularization parameters $\lm_l$ where the integral is absolutely convergent.
Moreover, Speer proves explicitly that such a domain of parameters $\lm_l$ is non-empty.

To prove this statement Speer uses the Feynman parametric representation of so analytically and dimensionally
regularized Feynman integral
\bea
\left(i\pi^{d/2} \right)^h
\frac{\Gamma(n+\sum \lm_i-h d/2)}{\prod_i \Gamma(1+\lm_i)}
&& \nn \\ &&  \hspace*{-47mm}
\times
\int_0^\infty \ldots\int_0^\infty \, \prod_i x_i^{\lm_i}
U^{n+\sum \lm_i-(h+1) d/2}
F^{h d/2-n-\sum \lm_i} 
\nn \\ &&  \hspace*{-47mm}
\times \delta\left( \sum x_i-1\right) \,
\dd x_1 \ldots \dd x_n\;,
\label{alpha-d-mod-lm}
\eea
and performs an analysis of convergence of~(\ref{alpha-d-mod-lm}) 
using sector decompositions. The goal of historically first sector decompositions~\cite{Hepp:1966eg,Speer:1975dc}
was to decompose a given parametric integral into sectors (subdomains) and then introduce new (sector)
variables in such a way that the singularities of the two basic polynomials 
$U$ and $F$ become factorized, i.e. in the sector variables they take the form of
a product of the sector variables raised to some powers times a function which is analytic and non-zero at
zero values of the sector variables. As a result, the analysis of convergence reduces to
power counting of the sector variables and each sector contribution of the analytically and dimensionally regularized
integral (\ref{alpha-d-mod-lm}) can be represented as a linear combination of products of typical factors
$1/(\ep+\sum'_i \lm_i)$ where $\ep=(4-d)/2$ and the sum is taken over a partial subset of parameters $\lm_i$.
After this, the singularities with respect to the regularization parameters are made manifest
and it becomes clear that integral~(\ref{alpha-d-mod-lm}) is a meromorphic function.
Speer suggests ~\cite{Speer:1975dc} to analytically continue this function to the point where all the $\lm$-parameters are zero
and thereby define dimensionally regularized version of~(\ref{alpha-d-mod})
even if there is no domain of $\ep$ where ~(\ref{alpha-d-mod}) is convergent.

Both Hepp~\cite{Hepp:1966eg} and Speer~\cite{Speer:1975dc} sectors are introduced {\em globally}, i.e. once and forever.
In fact, the Speer sectors\footnote{A variant of the Speer sectors is described in \cite{Smirnov:2008aw}; it is implemented
in the code {\tt FIESTA} \cite{Smirnov:2015mct}.} 
correspond to one-particle-irreducible subgraphs and their infrared analogues.
These sector decompositions were successfully applied for proving various results on regularized and renormalized
Feynman integrals. 

Global sector decompositions for Feynman integrals with on-shell infrared divergences
and/or collinear divergences are unknown. Binoth and Heinrich were first to construct {\em recursive} 
sector decompositions \cite{Binoth:2000ps,Binoth:2003ak,Heinrich:2008si}. 
The first step in their procedure was to introduce the 
set of primary sectors corresponding to the set of the lines of
a given graph, $\Delta_l = \left\{(x_1,\ldots,x_n) \left. \right| x_i\leq x_l,\; i\neq l\right\}$,
the sector variables are introduced by $x_i =  y_i x_l, \; i\neq l$.
The integration over $x_l$ is then taken due to the delta function 
in the integrand and one arrives at an integral over unit hypercube over $y_i$.

After primary sectors are introduced each sector integral
obtained is further decomposed into next sectors, according to some rule ({\em strategy}), 
and so on, until a desired factorization of the integrand in each resulting sector 
is achieved, i.e. it takes the form
\be
\int_0^1\ldots\int_0^1 f(y_1,\ldots,y_{n-1};\ep)\prod_{i=1}^{n-1} y_i^{a_i+b_i\ep} \dd y_i\,.
\label{factorized}
\ee
Here $y_i$ are sector variables in a final sector and 
a function $f$ is analytic in a vicinity of $y_i=0$ and is also analytic in $\ep$. (Remember that the number of integrations
is $n-1$ because one of the integrations was taken due to the delta function.)
Let us emphasize that such a factorization has a similar form, both in the
case of Hepp and Speer sectors and in final sectors within some recursive strategy.

To make singularities in $\ep$ explicit, one applies pre-subtractions in $y_i$
at zero values, i.e. for each integration with negative integer $a_i$, one adds and subtracts
first terms of the Taylor expansion,
\bea
\int_0^1 y^{a+b\ep}g(y)
=\sum_{k=0}^{-1-a}\frac{g^{(k)}(0)}{k!(a+k+b\ep+1)}
&& \nn \\ &&  \hspace*{-43mm}
+\int_0^1 y^{a+b\ep}\left[g(t)-
\sum_{k=0}^{-1-a}\frac{g^{(k)}(0)}{k!}t^k
\right]\,.
\label{presubtractions}
\eea
Therefore,  when a terminating strategy is applied, a given dimensionally regularized Feynman integral
is represented as a linear combination of {\em convergent} parametric integrals
with coefficients which are analytic functions of $\ep$.
 
There are several public codes where various strategies of recursive 
sector decompositions are implemented \cite{Bogner:2007cr,Carter:2010hi,Borowka:2015mxa,Borowka:2017idc,Smirnov:2015mct}.
In the case, where the basic polynomial $F$ is positive, Bogner and Weinzierl~\cite{Bogner:2007cr} 
presented first examples of strategies which terminate, i.e. provide, after a finite number of steps, a desired factorization
(\ref{presubtractions}) of the integrand in each final sector. 
 
When recursive sector decompositions are applied in practice, using a code for numerical evaluation, 
one does not care that, generally, there is no domain of
parameter $\ep$ where initial integral (\ref{alpha-d-mod}) is convergent.
However, in the case of Euclidean external momenta, one could remember about
the Speer's definition~\cite{Speer:1975dc} and use it to prove that this naive way is right.
Indeed, starting from the analytically regularized parametric representation (\ref{alpha-d-mod-lm})
and using some terminating strategy one can arrive at a factorization in final sector
of the form (\ref{factorized}), where the exponents of the final sector variables $a_i+b_i\ep$
obtain an additional linear combination of parameters $\lm_i$. Then one can use
the same procedure of making explicit poles in the regularization parameters by
a generalization of (\ref{presubtractions}). As a result one can observe that starting from
the Speer's domain of parameters $\lm_i$ where the given parametric integral is convergent
one can continue analytically all the terms resulting from the sector decomposition
and the procedure of extracting poles just by setting all the $\lm_i$ to zero.
 
However, extensions of the Speer's prescription to situations with on-shell infrared divergences
and/or col\-linear divergences are not available.
We are now going to provide such an extension. To do this we will use the 
LP parametric representation~(\ref{L-P}), rather than (\ref{alpha-d-mod})
and introduce  an auxiliary analytic regularization, i.e. turn from~(\ref{L-P}) to
\bea
\frac{
\Gamma(d/2)}{\Gamma((h + 1) d/2 -n-\sum \lm_i)\prod_i \Gamma(1+\lm_i)}                   
&& \nn \\ &&  \hspace*{-50mm}
\times
\int_0^\infty \ldots\int_0^\infty \, P^{-\dl} \, \prod_i x_i^{\lm_i} \,
  \dd x_1 \ldots \dd x_n\;,
\label{L-P-lm}
\eea
where now we keep all the factors. Although $\dl=2-\ep$ and $\lm_i$ are, generally, considered as
complex parameters, we will later consider them real, for simplicity.

In the next section, we will first derive conditions of convergence of integral~(\ref{L-P})
and then conditions of convergence of integral~(\ref{L-P-lm}). We will prove that there exists a non-empty 
domain of $\lm_i$ where the integral is convergent. 
Then, similarly to how this was done by Speer for Feynman integrals at Euclidean external momenta~\cite{Speer:1975dc},
we will formulate a definition of integrals~(\ref{L-P}) at general $\dl=2-\ep$
which, in particular, gives a definition of dimensionally regularized Feynman integrals
with possible on-shell infrared and collinear divergences.
 
\section{Convergence of the LP representation} 
 
Let $\pi(S)$ be the projection of the set $S$ on the hyperplane $w_{n+1}=0$, let $\pi(\mathcal{N}_P)$ 
be the projection of $\mathcal{N}_P$ on the same hyperplane, and $\pi(\gamma)$ be the corresponding projections
of essential facets.
It turns out that it is reasonable to turn to a more general family of integrals~(\ref{L-P})
by assuming that $P$ is given by (\ref{polynomial}) where the set $S$ is a finite set of {\em rational}
numbers. The following proposition holds.

{\bf Proposition 1.} The integral~(\ref{L-P}) is convergent if and only if 
$A=\left(\frac{1}{\dl},\ldots,\frac{1}{\dl}\right)\in {\mathbb R}^{n}$ 
is inside $\pi(\mathcal{N}_P)$.
 
{\bf Proof.}
1) Let us begin with the necessary condition.
It is clear that
the convergence of integrals 
\[\int_0^\infty\ldots\int_0^\infty
\left(\sum_{w\in \pi(S)} c_{w}x_1^{w_1}\ldots x_n^{w_n} \right)^{-\dl}  d x_1\ldots d x_{n}\]
with positive $c_{w}$ follows from the convergence of the integral
\[\int_0^\infty\ldots\int_0^\infty
\left(\sum_{w\in \pi(S)} x_1^{w_1}\ldots x_n^{w_n} \right)^{-\dl}  d x_1\ldots d x_{n}\]
and vice versa.
In particular, this means that the integral  
$G(t)$ defined by~(\ref{L-P}) for any  $t>0$ and the integral $G(1)=\int_0^\infty\ldots\int_0^\infty
\left(P(x,1)\right)^{-\dl}  d x_1\ldots d x_{n}$ are both convergent or both divergent.
Let us introduce notation $\widetilde{P}(x) = P(x,1)$. 

Let us assume that the statement is not true, i.e. that the integral $G(1)$ is convergent but
the point $A$ is outside the interior of the polytope  $\pi(\mathcal{N}_P)$.
Let us, first, consider the case where $A$ is outside $\pi(\mathcal{N}_P)$.
Since  $\pi(\mathcal{N}_P)$ is a convex set, there exist a plane 
$p_1w_1+\ldots+p_n w_n+p_0=0$
such that $\pi(\mathcal{N}_P)$ and $A$ are on its opposite sides.
One can choose a plane such that all $p_i\neq 0$. 
Let $p_1w_1+\ldots+p_n w_n+p_0<0$ for all the points $w$ of the polytope and
let $p_1\frac{1}{\dl}+\ldots+p_n \frac{1}{\dl}+p_0>0$, or $p_1+\ldots +p_n>-\dl p_0$.
 
Let us turn to the new variables $x_i=y_i^{p_i}$ in the integral $G(1)$.
We obtain
\bea
G(1)=\prod\limits_{i=1}^{n}|p_i|\cdot\int_0^\infty\ldots\int_0^\infty \prod \limits_{i=1}^{n} y_i^{p_i-1}
&& \nn \\ &&  \hspace*{-53mm}
\times \left(\sum_{w\in \pi(S)}c_w y_1^{p_1w_1}\ldots y_n^{p_nw_n} \right)^{-\dl}  d y_1\ldots d y_{n}.
\eea
Let us turn to hyperspherical coordinates. The new integration variables are
$r\in[0,+\infty)$, $\alpha_1,\ldots, \alpha_{n-1}\in [0;\pi/2]$. 
To ensure the convergence of $G(1)$ we need convergence of the integral over the variable $r$, i.e.
\bea
\int_0^\infty \frac{r^{p_1+\ldots+p_n - n}\cdot r^{n-1}dr}{\big(\widetilde{P}(r, \alpha_1, \ldots, \alpha_{n-1})\big)^{\dl}}.
\eea
The polynomial $\widetilde{P}(r, \alpha_1, \ldots, \alpha_{n-1})$ consists of terms 

\noindent $r^{p_1w_1+\ldots+p_nw_n}$ 
with coefficients depending on $\sin \alpha_i$ and $\cos\alpha_i$, and these coefficients are almost everywhere positive.
Therefore, in  order to have convergence at $+\infty$, one should have 
\[\dl  \max\limits_{w\in\pi(S)} (p_1w_1+\ldots+p_nw_n)-(p_1+\ldots +p_n)>0.\]
Since for all  $w\in\pi(S)$ we have 
 $p_1w_1+\ldots+p_nw_n<-p_0$ и $p_1+\ldots +p_n>-\dl p_0$, 
the left-hand side of this inequality is negative and we come to a contradiction. 

Let us now consider the case, where the point $A$ is at the boarder of the set $\pi(\mathcal{N}_P)$. 
Since $G(1)$ is a continuous function of $\dl$ then the convergence of the integral
as some $\dl$ leads to the convergence in a sufficiently small vicinity, i.e. once can find
an external point $(\frac{1}{\mu},\ldots,\frac{1}{\mu})$ of the polytope $\pi(\mathcal{N}_P)$, where
the integral $\int_0^\infty\ldots\int_0^\infty
\left(\widetilde{P}(x)\right)^{-\mu}  \dd x_1\ldots \dd x_{n}$ is convergent so that we come to a contradiction. $\Box$

2) Let us turn to the sufficient condition.
Let $\mathcal{K}$  be the set of vertices of a convex polytope which lies inside $\pi(\mathcal{N}_P)$.
To prove the sufficient condition, let us, first, show that if the integral 
$\int_0^\infty\ldots\int_0^\infty
\left(\widetilde{P}(x)\right)^{-\dl}  \dd x_1\ldots \dd x_{n}$ is divergent then the integral
\[\int_0^\infty\ldots\int_0^\infty\left(\sum_{w\in \mathcal{K}} x_1^{w_1}\ldots x_n^{w_n}\right)^{-\dl}  \dd x_1\ldots \dd x_{n}\]
is also divergent.
 
Here are two simple properties following from the comparison criterion of integrals:
  
(a) Let $Q_1$ and $Q_2$ be polynomials with positive coefficients. If the integral
\[\int_0^\infty\ldots\int_0^\infty\left(Q_1(x)+Q_2(x)\right)^{-\dl}  \dd x_1\ldots \dd x_{n}\]
is divergent then the integrals \[\int_0^\infty\ldots\int_0^\infty\left(Q_i(x)\right)^{-\dl}  \dd x_1\ldots \dd x_{n}\]
are also divergent.
  
(b) If a polynomial $Q(x)$ with positive coefficients contains terms  
$x_1^{w_1}\ldots x_n^{w_n}$ and $x_1^{u_1} \ldots x_n^{u_n}$, 
then the convergence of the following two integrals is equivalent:
\[\int_0^\infty\ldots\int_0^\infty
\left(Q(x)\right)^{-\dl}  \dd x_1\ldots \dd x_{n}\]
and
\[\int_0^\infty\ldots\int_0^\infty
\left(Q(x)+x_1^{\beta_1}\ldots x_n^{\beta_n}\right)^{-\dl}  \dd x_1\ldots \dd x_{n}\,,\]
where $\beta_i = w_i+z(u_i-w_i)$, $z\in[0,1]$,

The property (a) is obvious. The property (b) follows from the following inequalities
\bea
x_1^{w_1}\ldots x_n^{w_n}+x_1^{u_1} \ldots x_n^{u_n}<x_1^{w_1}\ldots x_n^{w_n}
&& \nn \\ &&  \hspace*{-55mm}
+x_1^{u_1} \ldots x_n^{u_n}+x_1^{\beta_1}\ldots x_n^{\beta_n}
\nn \\ &&  \hspace*{-58mm}
=x_1^{w_1}\ldots x_n^{w_n}\big(1+x_1^{u_1-w_1} \ldots x_n^{u_n-w_n}
\nn \\ &&  \hspace*{-55mm}
+(x_1^{u_1-w_1} \ldots x_n^{u_n-w_n})^z\big)\,.
\label{lab1}
\eea
If $x_1^{u_1-w_1} \ldots x_n^{u_n-w_n}\leq 1$ then the right-hand side of (\ref{lab1}) is less or equal to
\bea
x_1^{w_1}\ldots x_n^{w_n}\big(2+x_1^{u_1-w_1} \ldots x_n^{u_n-w_n}\big)
&& \nn \\ &&  \hspace*{-30mm}
\leq 2(x_1^{w_1}\ldots x_n^{w_n}+x_1^{u_1} \ldots x_n^{u_n})\,.
\eea
If $x_1^{u_1-w_1} \ldots x_n^{u_n-w_n}\geq 1$, then the right-hand side of (\ref{lab1}) is less or equal to
\bea x_1^{w_1}\ldots x_n^{w_n}\big(1+2x_1^{u_1-w_1} \ldots x_n^{u_n-w_n}\big)
&& \nn \\ &&  \hspace*{-30mm}
\leq 2(x_1^{w_1}\ldots x_n^{w_n}+x_1^{u_1} \ldots x_n^{u_n})\,.
\eea
 
Let $\mathcal{B}$ be the set of vertices of $\pi(\mathcal{N}_P)$. 
Using (a) and the condition of divergence of the integral
\[\int_0^\infty\ldots\int_0^\infty
\left(\widetilde{P}(x)\right)^{-\dl}  \dd x_1\ldots \dd x_{n}\]
we obtain divergence of the integral 
\[\int_0^\infty\ldots\int_0^\infty
\left(\sum_{w\in \mathcal{B}} x_1^{w_1}\ldots x_n^{w_n}\right)^{-\dl}  \dd x_1\ldots \dd x_{n}\,.\]

Let us choose an arbitrary convex polytope inside $\pi(\mathcal{N}_P)$, with the set of vertices $\mathcal{K}$,
and consider various lines through pairs of vertices of this polytope.
Let us denote by $\mathcal{H}$ the set of points of intersection of these lines with the facets
$\pi(\mathcal{N}_P)$.
Applying then several times property (b) we obtain that the convergence of the integral
$\int_0^\infty\ldots\int_0^\infty
\left(\sum_{w\in \mathcal{B}} x_1^{w_1}\ldots x_n^{w_n}\right)^{-\dl}  \dd x_1\ldots \dd x_{n}$
is equivalent to the convergence of a similar integral with the sum over the set
$\mathcal{B}\cup\mathcal{H}$, and, therefore to the convergence of the integral with the sum
over the set $\mathcal{B}\cup\mathcal{H}\cup\mathcal{K}$.

Hence, the integral
\[\int_0^\infty\ldots\int_0^\infty
\left(\sum_{w\in \mathcal{B}\cup\mathcal{H}\cup\mathcal{K}} x_1^{w_1}\ldots x_n^{w_n}\right)^{-\dl}  \dd x_1\ldots \dd x_{n}\]
is divergent and, therefore, according to property (a), the integral with the sum over $\mathcal{K}$
is also divergent.
 
Now, let the point $A=(\frac{1}{\dl},\ldots,\frac{1}{\dl})$ 
belong to the interior of the polytope $\mathcal{N}_P$, and let integral (\ref{L-P}) be divergent. 
Let us choose an $n$-dimensional hypercube lying inside the polytope and containing the point $A$
such that its facets are parallel to the axes.
The set of the vertices of the hypercube is $\mathcal{K}$ and, according to the statements above,
the integral 
\[\int_0^\infty\ldots\int_0^\infty
\left(\sum_{w\in \mathcal{K}} x_1^{w_1}\ldots x_n^{w_n}\right)^{-\dl} \dd x_1\ldots \dd x_{n}\]
is divergent. On the other hand, since $\mathcal{K}$ are the vertices of the chosen hypercube,
there are positive rational $q$ and $l$ such that this integral can be represented as
\bea
\int_0^\infty\ldots\int_0^\infty\prod\limits_{i=1}^{n}
\left( x_i^{q}\left(1+x_i^{l}\right) \right)^{-\dl}  \dd x_i 
&& \nn \\ &&  \hspace*{-40mm}
=\left(\int_0^\infty
\left( x^{q}(1+x^{l}) \right)^{-\dl}  \dd x\right)^n
\nn \\
\nn \\ &&  \hspace*{-40mm}
=\left(\frac{1}{l}B\left(\frac{1-\dl q}{l}, \frac{\dl (q+l)-1}{l}\right)\right)^n\,.
\eea
Since the point  $A$ is inside the hypercube, we have
 $q<\frac{1}{\dl}<q+l$ and the integral is convergent so that we come to a contradiction. $\Box$
  
Suppose now that the condition of Proposition~1 does not hold, i.e. the point
$A=\left(\frac{1}{\dl},\ldots,\frac{1}{\dl}\right)$
is not inside $\pi(\mathcal{N}_P)$. Then we introduce a general analytic regularization
and turn to integral~(\ref{L-P-lm}). We have 

{\bf Proposition 2.}
The integral~(\ref{L-P-lm}) is convergent 
if the point $\left(\frac{1+\lm_1}{\dl},\ldots,\frac{1+\lm_n}{\dl}\right)\in {\mathbb R}^{n}$ 
is inside $\pi(\mathcal{N}_P)$.

{\bf Proof.}
The proposition can be proven by the change of variables
$x_i \to x_i^{1/(\lm_i+1)}$ in (\ref{L-P-lm}). We then obtain 
$1/\prod_{i=1}^n (1+\lm_i)$ times the following integral
\bea
\int_0^\infty \ldots\int_0^\infty \,
\bar{P}^{-\dl}  \dd x_1 \ldots \dd x_n\;,
\label{L-P-bar}
\eea
where
\bea
\bar{P}(x_1,\ldots,x_n,t)=\sum_{v\in \bar{S}} c_{v}x_1^{v_1}\ldots x_n^{v_n} t^{v_{n+1}}\;,
\label{polynomial-bar}
\eea
with $\bar{S}=\left\{(v_1,\ldots,v_n,v_{n+1}\left| \right. v_i=w_i/(1+\lm_i),\right.$

\noindent $\left.i=1,\ldots,n; v_{n+1}=w_{n+1} \right\}$.
Using the convex property of the polytopes $\mathcal{N}_P$ and $\mathcal{N}_{\bar{P}}$ we arrive at the desired statement.

The function $\bar{P}$ is no longer a polynomial but we assume this possibility in Proposition~1.
Now, it is clear that we can adjust parameters $\lm_i$ using a blowing-down or blowing up 
(with $-1<\lm_i<0$ or $\lm_i>1$) to provide convergence by putting
$\frac{1+\lm_i}{\dl}$ between the left and the right values of the $i$-th
coordinates of $\pi(\mathcal{N}_P)$. $\Box$
 
Let us formulate this statement
as an analogue of the Speer's theorem~\cite{Speer:1975dc}.
 
{\bf Corollary 1.} 
The integral~(\ref{L-P-lm}) is an analytic function of parameters $\lm_i$ in a non-empty domain.

This domain exists for any given $\dl\equiv 2-\ep$. Now we define the integral~(\ref{L-P})
as a function of $\ep$ as the analytic continuation of the integral~(\ref{L-P-lm}) from
the convergence domain of parameters $\lm_i$ to the point where all $\lm_i=0$ by referring
to sector decompositions in the same way as it was outlined in the previous section.

It suffices then to explain how sector decompositions can be introduced for the LP integrals.
If we are dealing with a Feynman integral, with Eq.~(\ref{Pol}), we turn to (\ref{alpha-d-mod})
so that we can apply standard terminating strategies. If this is a more general integral,
with a positive polynomial $P$ one can reduce it to integrals over unit hypercubes, for example,
by the following straightforward procedure. 
Make the variable change $x_i=y_i/(1-y_i)$ to arrive at an integral over a unit hypercube. 
In order to avoid singularities near $y_i=1$, decompose each integration over $y_i$ in two
parts: from $0$ to $1/2$ and $1/2$ to $1$ and change variables again in order to have
integrations over unit hypercubes. As a result, one arrives at integrals to which
terminating strategies~\cite{Bogner:2007cr} can be applied.

\section{Equivalence of the new and the old prescriptions}

Up to now, the code {\tt asy.m}~\cite{Pak:2010pt} included in {\tt FIESTA}~\cite{Smirnov:2015mct}
was based on prescriptions formulated in Section~2 but with the use of the representation~(\ref{alpha-d-mod})
and the corresponding product $U F$ of the two basic functions, rather than
with the use of~(\ref{L-P}). Let us prove that the two prescriptions are equivalent.

Let us keep in mind that the functions $U$ and $F$ are homogeneous in the variables $x_i$, with different
homogeneity degrees.

\textbf{Proposition 3.}
Let $U$ and $F$ be two homogeneous functions of the variables $x_i$ with different homogeneity degrees such that
the Newton polytope $\mathcal{N}_{U+F}$ for $U+F$ has dimension $n+1$. Equivalently, $\mathcal{N}_{U F}$ has dimension $n$.
Then there is a one-to-one correspondence between essential facets of $\mathcal{N}_{U+F}$ and essential facets of $\mathcal{N}_{U F}$.
This correspondence is obtained by the projection on the hyperplane orthogonal to the vector $\{1,\ldots,1,0\}$ 
which we will denote by $v_0$.

\textbf{Proof.}
Let $\Gm$ be an essential facet of $\mathcal{N}_{U+F}$. It has dimension $n$. 
Since $\mathcal{N}_{U}$ and $\mathcal{N}_{F}$ have dimension not greater than $n$ this means that if $\Gm$ 
does not intersect with one of them
it should contain the other Newton polytope whose dimension is $n$. Then, due to homogeneity, its normal vector is 
proportional to $v_0$ but this
cannot be the case for an essential facet. Hence $\Gm$ has a non-empty intersection with both Newton polytopes.

Let us analyze intersections $\Gm_U$ and $\Gm_F$ of the facet $\Gm$ with
$\mathcal{N}_{U}$ and $\mathcal{N}_{F}$, correspondingly.
The hyperplane generated by $\Gm$ has dimension $n$ and can be defined as a vector sum
of the hyperplane generated by $\Gm_{U}$, the hyperplane generated by $\Gm_{F}$
and some vector which connects a point of $\Gm_{U}$ and a point of $\Gm_{F}$.
Therefore, the vector sum of the hyperplane generated by $\Gm_{U}$ and
the hyperplane generated by $\Gm_{F}$ has dimension $n-1$.

Furthermore, both hyperplanes are orthogonal to the vector $r^{\Gm}$ and to the vector $v_0$,
therefore they are also orthogonal to $r^{\Gm}_0$, the projection of the vector $r^{\Gm}$ on the hyperplane orthogonal to $v_0$.

Now it suffices to show that $r^{\Gm}_0$ corresponds to a facet of $\mathcal{N}_{UF}$.
Indeed, the minimal values of scalar products of points of this polytope with the vector $r^{\Gm}_0$
is achieved from the pairwise sums of the points of the facets of $\Gm_U$ and $\Gm_F$.
The linear space spanned by these points can be generated by the vector sum of the hyperplanes spanned over the sets
$\Gm_{U}$ and $\Gm_{F}$ but we have just shown that this space has dimension $n-1$, i.e is a facet.

Now let us turn to the inverse statement.
Let $\Gm$ be a facet of $\mathcal{N}_{UF}$.
Its normal vector $r^{\Gm}$ is orthogonal to $v_0$.
Let us consider the sets 
$\nu_U$ and $\nu_F$ consisting of points with the minimal scalar product with $r^{\Gm}_0$
of $\mathcal{N}_{U}$ and $\mathcal{N}_{F}$, respectively.

The sum of the hyperplanes spanned on $\nu_U$ and $\nu_F$
coincides with the hyperplane spanned on $\Gm$. Therefore, it has dimension $n-1$.

Let the scalar product of $r^{\Gm}_0$ and the points of $\nu_U$ be $u_0$
and the scalar product of $r^{\Gm}_0$ and the points of $\nu_F$ be $f_0$.
Furthermore, let the scalar product of $v_0$ and points of $\nu_U$ be $u_1$
and the scalar product of $v_0$ and points of $\nu_F$ be $f_1$
The fact that the scalar product is fixed follows from the homogeneity, and we know than $u_1 \neq f_1$.

Let us find such a vector $r = r^{\Gm}_0 + \{\alpha,\ldots,\alpha,0\}$
that its scalar product with points of $\nu_U$ and $\nu_F$ is the same.
To do this we solve the equation $u_0 + x u_1 = f_0 + \alpha f_1$, so that $\alpha = (f_0-u_0) / (u_1 - f_1)$.

Let $\Gm$ be the face of $\mathcal{N}_{U F}$ spanned over points having the minimal product with $r$.
The hyperplane spanned over $\Gm$ is the sum of hyperplanes spanned over
$\nu_U$ and $\nu_F$ and some vector connecting a point
of $\nu_U$ and a point of $\nu_F$.
The dimension of the sum of first two hyperplanes is $n-1$ however the connecting vector does not
belong to this sum since it is not orthogonal to $v_0$. Therefore
$\Gm$ is a facet of $\mathcal{N}_{U F}$ and $r = r^{\Gm}$. $\Box$

\section{The leading order}

Let us, first, assume that the conditions of Proposition~1 hold. We have

{\bf Proposition 4.} 
If the point $A=\left(\frac{1}{\dl},\ldots,\frac{1}{\dl}\right)\in {\mathbb R}^{n}$ 
is inside  $\pi(\Gamma)$ for some facet $\Gm$ then the leading asymptotics of the integral (\ref{L-P})
is given by Eq.~(\ref{L-P-LO}), i.e.
\bea
G(t,\ep)&\sim& 
M_{\Gm} G(t,\ep) \equiv t^{-L(\Gm)\dl+\sum_i r^{\Gm}_i} \nn \\
\nn \\ &&  \hspace*{-15mm}
\times
\int_0^\infty
\ldots\int_0^\infty\!\!\left( \sum_{w\in\Gamma\cap S} c_{w}y_1^{w_1}\ldots y_n^{w_n} \right)^{-\dl}\!\!\dd y_1\ldots \dd y_n 
\label{Prop4}
\eea
when $t\rightarrow +0$, where $r_i$ and $L(\Gm)$ are defined in Section~2.
 
{\bf Proof.}
Let us observe that for $w\in \Gamma$ we have $w_{n+1}=-\sum\limits_{i=1}^n r_i^{\Gamma}w_i +L(\Gamma)$, and,
since $\mathcal{N}_P$ is a convex set, 
we have
$w_{n+1}>-\sum\limits_{i=1}^n r_i^{\Gamma}w_i +L(\Gamma)$, for
$w\in S\setminus\Gamma$, i.e.
$w_{n+1}=-\sum\limits_{i=1}^n r_i^{\Gamma}w_i +L(\Gamma)+\kappa_{w,\Gamma}$, where $\kappa_{w,\Gamma}>0$. 
If we change variables $x_i=t^{r_{i}^{\Gamma}}\cdot y_i$ in the integral (\ref{L-P})
we obtain
\bea
\mathcal{F}(t)&=&t^{-L(\Gm)\dl+\sum_{i=1}^n r^{\Gm}_i} 
\nn \\ &\times&
\int\ldots\int
(\Phi (y,t))^{-\dl}
\dd y_1\ldots \dd y_n \,,
\label{Prop4a}
\eea
where
\bea
\Phi (y,t)&=&\phi(y) +\sum_{ 
S\setminus\Gamma} c_{w} y_1^{w_1}\ldots y_n^{w_n} t^{\kappa_{w,\Gamma}} \;,
\nn \\
\phi(y)&=&\sum_{ \Gamma
\cap S} c_{w} y_1^{w_1}\ldots y_n^{w_n}
\;.
\label{phis}
\eea    
 
Let us observe that
$\Phi^{-\dl}(y,t)$ is a positive continuous function of $n+1$ variables which is non-decreasing at
any fixed $y$ with respect to $t$ when  $t\rightarrow +0$. Moreover, we have
$ \Phi^{-\dl}(y,t)\rightarrow\phi^{-\dl}(y)$, and the integral
$\int\limits_{0}^{\infty}\ldots\int\limits_{0}^{\infty}
\phi^{-\dl}(y)dy$ is convergent because the point
$A=(\frac{1}{\dl},\ldots,\frac{1}{\dl})$ belongs to the interior of  $\pi(\Gamma)$ (according to
Proposition~1).
 
Then, using a theorem about the continuity of an integral depending on a parameter,  at
$t\rightarrow +0$ 
we obtain
\bea\int\limits_{0}^{\infty}\ldots\int\limits_{0}^{\infty}
\Phi^{-\dl}(y,t)dy\rightarrow 
\int\limits_{0}^{\infty}\ldots\int\limits_{0}^{\infty}
\phi^{-\dl}(y)dy
\eea
so that we arrive at (\ref{Prop4}). $\Box$
 

If the condition of Proposition~1 does not hold we can use Proposition~2 and 
adjust an analytic regularization to provide convergence. Let $\Gm$ be an essential facet.
Then, like in the proof of Proposition~2, we can adjust parameters $\lm_i$ by putting
$\frac{1+\lm_i}{\dl}$ between the left and the right values of the $i$-th
coordinates of $\pi(\Gm)$. After this, we can follow the same arguments as in the proof of
Proposition~3 and obtain the following generalized version.

{\bf Proposition~5.} 
Let $G(t;\ep)$ be integral (\ref{L-P})
with a polynomial (\ref{polynomial}) and let $\Gm$ be an essential facet.
Then one can adjust analytic regularization parameters $\lm_i$,  i.e. to turn to the integral
$G(t;\ep;\lm_1,\ldots,\lm_n)$ defined by
(\ref{L-P-lm}), by satisfying the condition
$\left(\frac{1+\lm_1}{\dl},\ldots,\frac{1+\lm_n}{\dl}\right)\in \pi(\Gamma)$, 
so that the contribution of this facet to the expansion of 
$G(t;\ep;\lm_1,\ldots,\lm_n)$ will be leading and  have the form (up to a coefficient independent of $t$)
\bea
t^{-L(\Gm)\dl+\sum_{i=1}^n (\lm_i+1) r^{\Gm}_i}  \int_0^\infty
\ldots\int_0^\infty\prod_{i=1}^n y_i^{\lm_i} 
&& \nn \\ &&  \hspace*{-55mm}
\times
\left( \sum_{w\in\Gamma\cap S} c_{w}y_1^{w_1}\ldots y_n^{w_n} \right)^{-\dl}\, 
\dd y_1\ldots \dd y_n
\label{Prop5}
\eea
when $t\rightarrow +0$.
 
Let us emphasize that the projection $\pi(\mathcal{N}_P)$ of the Newton polytope can be covered
by the corresponding projections $\pi(\gm)$ of essential facets.
The intersection of any pair $\pi(\gm_1)$ and $\pi(\gm_2)$ of projections of the facets has dimension less than $n$.
Therefore, the contribution of one of the facets can be made leading by
adjusting analytic regularization parameters.
We can refer again to sector decompositions in order to prove that the contribution
of each facet is a meromorphic function of parameters $\ep$ and $\lm_i$ so that then we can expand
a result for this contribution in the limit of small $\lm_i$ up to a finite part in $\lm_i$
keeping possible singular terms in $\lm_i$, and then expand in $\ep$ at $\ep\to 0$.

We can use this procedure as an unambiguous definition of the leading contribution of a given facet,
i.e. we can clarify the prescriptions in Section~2 and define it as the expanded analytic continuation
of the contribution described in Proposition~5 in the two successive limits, $\lm_i\to 0$
and $\ep \to 0$. However, it is necessary to specify how the limit $\lm_i\to 0$ is taken.
At least two practical variants were in use: (1) take the limits $\lm_i\to 0$ for $i=1,2,\ldots$,
or in some other fixed order, keeping expansion up to $\lm_i^0$; (2) choose, $\lm_i=p_i \lm_1$, $i=2,3,\ldots$, 
where $p_i$ is the i-th prime number and
then take the limit $\lm_1\to 0$. The second variant was systematically used, in particular, in Refs.~\cite{Henn:2014lfa,Caola:2014lpa}.
In both cases, the definitions depends on the order of parameters $\lm_i$ but
final results for the whole expansion should be independent of this choice if the initial integral is convergent
at $\lm_i=0$.

Now, we can compose the sum
\bea
\sum_i M_i^0 G(t,\ep,\lm_1,\ldots,\lm_n)\,,
\label{LO}
\eea
where each term is convergent in the corresponding domain of $\lm_i$ and where it is the leading term
of the whole expansion. Let us refer again to theorems on sector decompositions~\cite{Bogner:2007cr} which
make manifest the analytic structure with respect to the regularization parameters $(\ep,\lm_1,\ldots,\lm_n)$
in order to claim that each term can be continued analytically to a sufficiently small vicinity of the point $(\ep,0,\ldots,0)$.
Let us assume that, at a given $\ep$, the initial integral is analytic. (This can be checked with sector decompositions.)
In particular, this happens if at this $\ep$, the initial integral is finite. 
Then it turns out that the limit of (\ref{LO}) at $\lm_i\to 0$ gives the leading order terms 
in accordance with our main conjecture so that it looks like we have justified it.
However, here we implied that the operations of expansion and analytic continuation commute.
We believe that this is indeed the case and hope that this property can be proven.
 
It is clear that one has to choose the same way of taking the limit $\lm_i\to 0$
for all the facets. Possible individual singularities in $\lm_i$ should cancel in the sum of contributions of
different facets. Then $\lm_i\to 0$ and we are left with expansion in $\ep$. 
Of course, the order of contributions to the expansion is measured in powers of $t$ when the limit $\lm_i\to 0$
is already taken. The true leading order of the expansion is given by a sum of contributions
of some essential facets which can be called leading.

\section{General order for one essential facet} 
 
Let us consider a simple situation with  one essential facet. For Feynman integrals, this can be,
for example, an expansion in the small momentum limit, where a given Feynman graph has no massless
thresholds. Then one can refer to general analytic properties of Feynman amplitudes
and claim that the Feynman integral is analytic up to the first threshold so that
if can be expanded in a Taylor series at zero external momenta. Of course, there is
only one essential facet in the corresponding Newton polytope associated with
the polynomial $P$ in (\ref{L-P}) and the limit looks trivial.
However, our goal is an integral with an arbitrary polynomials with positive coefficients, so that
the situation with one essential facet should not be qualified as trivial.
We have the following 
 
{\bf Proposition 6.} If there is only one essential facet $\Gm$ in the Newton polytope then
\bea 
\mathcal{F}(t)\sim
&& \nn \\ &&  \hspace*{-13mm}
\int_0^\infty\ldots \int_0^\infty \left[M_{\Gamma} \left(P(x_1,\ldots,x_n,t)\right)^{-\dl}\right] \dd x_1\ldots \dd x_n
\label{1fgenorder}
\eea
when $t\rightarrow +0$.

{\bf Proof.} 
Let us start from Eq.~(\ref{Prop4a}). The second term in the brackets tends to zero at 
$t\rightarrow +0$, so that one can obtain a series in powers of $t$ by expanding this expression
with respect to the second term, according to the prescriptions formulated in Section~2. 
This is, generally, not a Taylor expansion. Rather, this is
an expansion in powers of $t^{1/q}$ where $q$ is the
least common multiple of the rationals $\kappa_{w,\Gamma}$.

The coefficients at powers of $t$ in the resulting sum in the integrand have the following form (up to constants):
\bea
E(y_1,\ldots,y_n)=\prod\limits_{i=1}^{n} y_i^{\sum \limits_{j=1}^{m} u_i^j k_j} 
&& \nn \\ &&  \hspace*{-23mm}
\times
\Big(\sum_{ \Gamma
\cap S} c_{w} y_1^{w_1}\ldots y_n^{w_n} \Big)^{-m-\dl}\,,
\label{Ef}
\eea
where $m=0,1,2,\ldots$ are powers of the Taylor expansion with respect to the second term, 
$k_j$ are non-negative integers, 
$\sum \limits_{j=1}^{m} k_j=m$, 
and the points $u^j=(u_1^j,\ldots,u_n^j)$, $j=1,\ldots,m$ belong to the projection of
$\pi(S\setminus \Gamma)$ on the plane $w_{n+1}=0$.

Let us define $\widetilde{u}_i= \sum \limits_{j=1}^{m} u_i^j k_j$. 
Taking into account the convex property of the set $\pi(\mathcal{N}_P)$,
the property of $k_j$ and the fact that there is only one essential facet $\Gamma$, we can conclude
that the point $\frac{1}{m}(\widetilde{u}_1,\ldots, \widetilde{u}_n)$ is an internal point of $\pi(\mathcal{N}_P)$.
Let us prove, using Proposition~2, that the convergence property of the integral (\ref{Ef}) of $E(y_1,\ldots,y_n)$  is equivalent to
the condition that the point $A=\frac{1}{m+\dl}(\widetilde{u}_1+1,\ldots, \widetilde{u}_n+1)$  is inside $\pi(\mathcal{N}_P)$. 
Let us assume that this is not true, i.e. $A$ is not an internal point of $\pi(\mathcal{N}_P)$.
Then there should exist a hyperplane $\sum\limits_{i=1}^n p_iw_i+p_0=0$ such that $\pi(\mathcal{N}_P)$ 
and $A$ belong to the different sides from this hyperplane, or on this hyperplane.
 
We have the following four conditions
\begin{enumerate}
\item 
The inequality $\sum\limits_{i=1}^n p_iw_i+p_0\leq 0$ holds for $w\in\pi(\mathcal{N}_P)$;
\item 
The relation  $\frac{1}{\dl+m}\sum\limits_{i=1}^n p_i(\widetilde{u}_i+1)+p_0\geq 0$
holds for the point $A$;
\item 
The condition of convergence of the initial integral is
$\frac{1}{\dl}\sum\limits_{i=1}^n c_i+c_0<0$; 
\item  
The relation $\sum\limits_{i=1}^n p_i(\frac{1}{m}\widetilde{u}_i-\widetilde{w}_i)<0$
holds for some point  $\widetilde{w}\in\pi(\mathcal{N}_P)$ 
because
$\frac{1}{m}(\widetilde{u}_1,\ldots, \widetilde{u}_n)$ is an internal point of the convex set 
$\pi(\mathcal{N}_P)$.
\end{enumerate}

Using these four conditions we arrive at the following chain of inequalities:
\bea
0&\stackrel{2.}\leq&\sum\limits_{i=1}^n p_i(\widetilde{u}_i+1)+p_0(\dl+m)
\nn \\
&=&\sum\limits_{i=1}^n p_i\widetilde{u}_i+\sum\limits_{i=1}^n p_i +p_0\dl+p_0m
\nn \\
&\stackrel{3.}<&\sum\limits_{i=1}^n p_i\widetilde{u}_i+p_0m\stackrel{4.}<m\sum\limits_{i=1}^n p_i\widetilde{w_i}+p_0m
\nn \\
&=&m\Big( \sum\limits_{i=1}^n p_i\widetilde{w}_i+p_0 \Big)\stackrel{1.}\leq 0\,.
\eea
 
As as result, we come to a contradiction so that the integrals of  $E(y_1,\ldots,y_n)$
are convergent. This means that (\ref{1fgenorder}) is true. One can represent this expansion
by introducing an auxiliary parameter, $\rho$, into the second term in the square brackets in 
Eq.~(\ref{Prop4a}) and perform an expansion in $\rho$ at $\rho \to 0$ and setting $\rho=1$ in the end. 
$\Box$

\section{Summary}
 
We advocated the Lee--Pomeransky representation~(\ref{L-P}) \cite{Lee:2013hzt} as a means to describe and to prove
expansion by regions. Starting from the prescriptions
of expansion by regions which were earlier implemented in the code {\tt asy.m}~\cite{Pak:2010pt} 
included in {\tt FIESTA}~\cite{Smirnov:2015mct} and now reformulated with the use of 
the LP representation~(\ref{L-P}) we clarified these prescriptions and made first steps towards
their justification. 
\begin{itemize}
\item We performed an analysis of convergence of the LP representation, proved a generalization of
the Speer's theorem for integrals~(\ref{L-P}) and presented a general definition of dimensionally regularized
integrals~(\ref{L-P}).
\item We presented a direct proof of equivalence of expansion by regions for Feynman integrals
based on the standard Feynman parametric representation~(\ref{alpha-d-mod}) and the LP representation~(\ref{L-P}).
This change is now implemented in {\tt FIESTA}~\cite{Smirnov:2015mct} so that revealing regions is now performed
in a much more effective way just because the degree of polynomial $P=U+F$ in (\ref{L-P}) is less than
the degree of the product of the polynomials $U F$.
\item We proved our prescriptions for the contribution of the leading order for each essential facet.
\item We proved our prescriptions in the general order in the simple situation with
one essential facet.
\end{itemize}
 
Let us emphasize that the use of an auxiliary analytic regularization is very natural 
to explicitly define dimensionally regularized
integrals~(\ref{L-P}). However, its use for the definition of individual contributions of
facets to the expansion in a given limit is even more important because, otherwise, these
terms can be ill-defined.
 
We believe that the commutativity of the expansion procedure with the operation of
analytic continuation with respect to the regularization parameter can be proven so that
this will give a justification of the prescriptions at least in the leading order of expansion.
Another possible scenario would be to prove the prescriptions
in a general order of expansion by constructing
a remainder with the help of the operator $\prod_i(1-M_i^{n_i})$ with appropriately adjusted
subtraction degrees $n_i$.
The problem would be divided into two parts: justifying the necessary asymptotic estimate of the remainder
where an auxiliary analytic regularization is not needed and 
obtaining terms of the corresponding expansion where, generally,
an analytic regularization is necessary.

\begin{acknowledgements}
The work was supported by RFBR, grant 17-02-00175A.
We are grateful to Roman Lee and Alexey Pak for stimulating discussions. 
\end{acknowledgements}



\end{document}